\documentclass[aps,pre,twocolumn,showpacs,floatfix,superscriptaddress]{revtex4}
\usepackage{graphicx}

\usepackage{amssymb}
\usepackage{amsmath}

\newcommand{\caH}{{\mathcal H}}

\begin{document}

\title{A graph theoretical analysis of the energy landscape of model polymers}

\author{Marco Baiesi}
\affiliation{Dipartimento di Fisica and Centro
interdipartimentale per lo Studio delle Dinamiche
Complesse (CSDC), Universit\`a di Firenze,
via G.\ Sansone 1, 50019 Sesto Fiorentino, Italy}
\altaffiliation{Institute of Theoretical Physics, 
K.\ U.\ Leuven, Celestijnenlaan 200D, Leuven, Belgium}
\author{Lorenzo Bongini}
\email{bongini@fi.infn.it}
\affiliation{Dipartimento di Fisica and Centro
interdipartimentale per lo Studio delle Dinamiche
Complesse (CSDC), Universit\`a di Firenze,
via G.\ Sansone 1, 50019 Sesto Fiorentino, Italy}
\author{Lapo Casetti}
\affiliation{Dipartimento di Fisica and Centro
interdipartimentale per lo Studio delle Dinamiche
Complesse (CSDC), Universit\`a di Firenze,
via G.\ Sansone 1, 50019 Sesto Fiorentino, Italy}
\affiliation{Istituto Nazionale di Fisica Nucleare (INFN), sezione di
Firenze, via G.\ Sansone 1, 50019 Sesto Fiorentino, Italy}
\author{Lorenzo Tattini}
\affiliation{Dipartimento di Chimica, Universit\`a di Firenze,
via della Lastruccia 3, 50019 Sesto Fiorentino, Italy}

%%%%%%%%%%%%%%%%%%%%%%%%%%%%%%%%%%%%%%%%%%%%%%%%%%%%%%%%%%%%%%%%%%%%%%%%%%%
\begin{abstract}

In systems characterized by a rough potential energy landscape, local energetic minima and saddles 
define a network of metastable states whose topology strongly influences the
dynamics. Changes in temperature, causing the merging and splitting of metastable
states, have non trivial effects on such networks and must be taken into account.
We do this by means of a recently proposed renormalization procedure. 
This method is applied to analyze the topology of the network of
metastable states for different polypeptidic sequences in a minimalistic polymer model. A smaller
spectral dimension emerges as a hallmark of stability of the global energy minimum and highlights a 
non-obvious link between dynamic and thermodynamic properties.

\end{abstract}
%%%%%%%%%%%%%%%%%%%%%%%%%%%%%%%%%%%%%%%%%%%%%%%%%%%%%%%%%%%%%%%%%%%%%%%%%%

\maketitle

%%%%%%%%%%%%%%%%%%%%%%%%%%%%%%%%%%%%%%%%%%%%%%%%%%%%%%%%%%%%%%%%%%%%%%%%%%%%%%%%%%%%%%%%%%%%%%%%%%%%%%%%%%%%%
%%%%%%%%%%%%%%%%%%%%%%%%%%%%%%%%%%%%%%%%%%%%%%%%%%%%%%%%%%%%%%%%%%%%%%%%%%%%%%%%%%%%%%%%%%%%%%%%%%%%%%%%%%%%%
%%%%%%%%%%%%%%%%%%%%%%%%%%%%%%%%%%%%%%%%%%%%%%%%%%%%%%%%%%%%%%%%%%%%%%%%%%%%%%%%%%%%%%%%%%%%%%%%%%%%%%%%%%%%%

\section{Introduction}
\label{uno}

The dynamical behavior of polymers in the collapsed phase has been a field of active research since the
concepts of $\Theta$ solvent and collapse transition were introduced by Flory in the early fifties of the XX century
\cite{flory}. In the last decades many statistical mechanical analyses of polymer models have been carried
out \cite{review}, studying configurational plasticity and dynamics, and their dependence on the microscopic
details of the  physical interactions between monomers and with the solvent. However, the inherent
frustration of these models often hinders analytical approaches, and even the investigations of very simple
polymer models still have to be undertaken on a numerical basis. This fundamental difficulty also affects the
study of an important class of heteropolymers, namely proteins. According to their sequence these molecules
fold into a unique tridimensional structure, the native configuration, which corresponds to the minimum of
the potential  energy. Natural proteins show a remarkable variety of folding behaviors. Folding times might
span a couple of orders of magnitude even for proteins of comparable size \cite{rates} and folding kinetics
of very similar proteins might change from two-stage to more complex, multistage transition schemes
\cite{twostate}. Moreover, in the last twenty years, the development of combinatorial methods of protein
synthesis has allowed the experimental verification of an old paradigm of protein science: random sequence
polypeptides very rarely fold \cite{random}. It is, however, still not clear how exactly proteins differ from
random heteropolymers and how these in turn differ from a homopolymer.

Minimalistic models might be useful in attacking such basic questions because they can be investigated
more easily than complex and more realistic models. Since the proposal
of the lattice HP model \cite{dill} several protein models have proven capable of qualitatively
reproducing the main thermodynamic features of the folding process, such as the folding and
$\Theta$-transition temperatures, while taking into account the effect of
different primary sequences \cite{models}. Most of these models exhibit a funneled
energy landscape characterized by the presence of many competing local minima of the potential
energy \cite{wales}, in strict analogy with structural glasses \cite{wales,parisi,angelani}.
In this framework the folding properties of a given protein sequence are often depicted as arising
from the interplay between funnel steepness
(the global bias toward the native configuration) and the landscape roughness (the number and
average depth of energy minima). More precisely each protein is characterized by a well defined
temperature range in which it manages to attain its native structure. When this range is
sufficiently large the corresponding sequence can be defined  a ``good folder''.
The same definition is often applied also to fast folding sequences, in analogy to the short
folding times that generally characterize real proteins. The relations between the folding properties 
of proteins and the topography of their energy landscape have been investigated at length leading to the 
evidence of complex kinetic behaviors \cite{wang1,wang2} and to the proposal of several criteria 
for the identification of fast folders using equilibrium indicators \cite{veit,tlp}.

A very promising method for analyzing the topography of energy landscapes is representing them
as a network.
Dynamic trajectories on rough energy landscapes typically exhibit a separation of time scales
in the sampling of the available configurational space. The first-order saddles of the
potential that connect the basins of attraction of different minima also represent kinetic
bottlenecks for the system and usually induce a partitioning of the configuration space in a
finite set of metastable states, each characterized by a large escape time and a fast local
diffusivity.
In this paper we will stress that a metastable state in a system at finite temperature
can either correspond to the basin of attraction of a single minimum of the potential energy or,
more generally, to a collection of different minima linked by
low-energy saddles.
Under these conditions the long-term dynamics of the system consists of  a series
of activated jumps between different metastable states.
Trajectories can therefore be described
by a master equation which depends only on the transition rates between different metastable states
\cite{vankampen}. In this context the investigation of the statistical and topological
properties of the graph representing the network of connections between metastable states (NMS)
provides a tool for describing the structural organization of the landscape and its influence
on the folding dynamics.

Preliminary work on a two-dimensional toy model \cite{grafi2D} has shown that there are quantitative
differences in the topography of the energy landscape of heteropolymers and homopolymers.
Differences  are found also between fast folding and slow folding heteropolymers. 
However, although the model used in \cite{grafi2D} correctly shows all the distinctive thermodynamic 
phases expressed by random heteropolymers and proteins, its two-dimensional character raises 
serious concerns about its capability of reproducing the conformational flexibility of real 
polymers.
One of the goals of this work is, therefore, to explore the topography of the energy landscape of different polymers 
and heteropolymers using a more realistic, although still minimalistic, representation.

In this paper we investigate the topology of the NMS of the energy landscape of different
sequences in a minimalistic three-dimensional off-lattice polymer model. In Sec.~\ref{sec1} and
Sec.~\ref{sec2} we describe the model and the technique employed to sample the relevant fixed points
of the energy landscape. In Sec.~\ref{sec3} we present a renormalization procedure suitable to
group the basins of attraction of the existing minima of the potential energy into temperature
dependent metastable states. In Sec.~\ref{sec4} we briefly discuss the thermodynamic properties
of the systems while in the following sections we analyze the topology of
their NMS. More precisely, in Sec.~\ref{sec5} we detail the change in topology for increasing
chain length in hydrophobic homopolymers and in Sec.~\ref{sec6} we investigate the topological
differences between the NMS of two heteropolymeric sequences characterized by very different
stabilities of the native structure.

%%%%%%%%%%%%%%%%%%%%%%%%%%%%%%%%%%%%%%%%%%%%%%%%%%%%%%%%%%%%%%%%%%%%%%%%%%%%%%%%%%%%%%%%%%%%%%%%%%%%%%%%%%%%%
%%%%%%%%%%%%%%%%%%%%%%%%%%%%%%%%%%%%%%%%%%%%%%%%%%%%%%%%%%%%%%%%%%%%%%%%%%%%%%%%%%%%%%%%%%%%%%%%%%%%%%%%%%%%%
%%%%%%%%%%%%%%%%%%%%%%%%%%%%%%%%%%%%%%%%%%%%%%%%%%%%%%%%%%%%%%%%%%%%%%%%%%%%%%%%%%%%%%%%%%%%%%%%%%%%%%%%%%%%%

\section{The model}
\label{sec1}

We consider an off-lattice coarse-grained model for short peptides that has been
recently studied by Clementi and coworkers \cite{clementi00,clementi05,mossa07}.
The model describes a linear molecule with $N$ residues, each one representing the
position $\vec{x}_i$ of a $C_\alpha$ atom.
The bond vectors $\vec{r}_i$ are $\vec{x}_{i+1} - \vec{x}_{i}$, with length
$r_i=|\vec{r}_i|$.
Following the notation of Ref.~\cite{mossa07}, we define also the distances
$r_{i,j}=|\vec{x}_i - \vec{x}_j|$. An angle between subsequent bonds $\vec{r}_i$ and $\vec{r}_{i+1}$
is denoted by $\theta_i$, while $\varphi_i$ is the dihedral angle formed by the
residues $i,i+1,i+2,i+3$.
The potential energy has already been discussed in appendices of
Refs.~\cite{clementi00,mossa07}. It is composed of the following  contributions:
\begin{equation}
V =  V_{\rm bond} + V_{\rm ang} + V_{\rm dih} + V_{\rm LJ}\;.
\end{equation}
One has a bond term
\begin{equation}
V_{\rm bond} = k_\mathrm{r}\sum_{i=1}^{N-1}(r_i-r^{(0)}_i)2 \;,
\end{equation}
an angular term
\begin{equation}
V_{\rm ang} = k_\theta\sum_{i=1}^{N-2}(\theta_i-\theta_i^{(0)})2 \;,
\end{equation}
a dihedral term
\begin{equation}
\begin{split}
V_{\rm dih} = \sum_{i=1}^{N-3} \bigg[\,& k_\varphi^{(1)}(1-\cos(\varphi_i-\varphi_i^{(0)})) +\\
&k_\varphi^{(2)}(1-\cos 3(\varphi_i-\varphi_i^{(0)})) \bigg] \;,
\end{split}
\end{equation}
and a Lennard-Jones term
 \begin{eqnarray}
\begin{split}
V_{\rm LJ} =\, & \epsilon_1{\sum_{(i,j)\in C}}^{\prime}\left[ 5\left(\frac{\sigma_{i,j}}{r_{i,j}}\right)^{12} -
   6\left(\frac{\sigma_{i,j}}{r_{i,j}}\right)^{10}  \right]+\\
   & \epsilon_2{\sum_{(i,j)\notin C}}^{\prime} \left(\frac{\sigma_0}{r_{i,j}}\right)^{12} \;.
\end{split}
\end{eqnarray}

The notation $\sum^{\prime}$ refers to a sum over pairs $i,j$ with $j-i\ge 4$, i.e., over pairs of residues 
that are not consecutive to each other on the chain.
The choice of the contact map $C$, of the constants $r_i^{(0)},\theta_i^{(0)},\varphi_i^{(0)}$,
and of the distances $\sigma_{i,j}$, determines the kind of the peptide; the values chosen for these constants as well as for the remaining ones are reported below.

In this paper we aim at understanding relations between the topography of the energy landscape
of a polymer and the topology of its NMS. We also investigate how these features are influenced by the 
conformational stability and the size of the system. 
We have therefore analyzed three cases: two heteropolymers with $N=12$ monomers, characterized by
\begin{itemize}
\item[(a)] a highly stable native configuration (which from now on we will call the {\em stable folder})
\item[(b)] a highly unstable native configuration (which from now on we will call the {\em unstable folder})
\end{itemize}
and
\begin{itemize}
\item[(c)] a hydrophobic homopolymer with $4 \le N \le 12$.
\end{itemize}
The latter has been considered because it represents a system amenable to be studied at different sizes 
without introducing non-obvious sequence dependent effect, as one would do with heteropolymers.
As a stable folder we chose a $\alpha$-helix studied in~\cite{mossa07}. Analogously
to real helices, this system is characterized by a strong energetic bias toward the native configuration,
granted by a Go-like potential with $C_{i,j}=1$ only for $j=i+4$
that mimics the hydrogen bonding responsible for helix
stabilization. On the contrary the unstable sequence was chosen as the portion from residue
$i=8$ to residue $17$ that forms a beta-sheet in the model considered in \cite{clementi00}.
This segment is stabilized by interactions with residues not included in our selection and is
therefore highly unstable and substantially unstructured when isolated.

Finally for homopolymers we used
$C_{i,j}=1$ (still with $j\ge i+4$) to induce a generic attraction  between residues, and parameters
have been homogeneously fixed for all residues to $r_i^{(0)}=3.8$, $\theta_i^{(0)}=1.6$,
$\varphi_i^{(0)}=0.88$,  $\sigma_{i,j}=6.30$, $\sigma_0=3$, $\epsilon_1=10$,
$\epsilon_2=5\epsilon_1$.
In all cases (a)-(c), the dihedral potential has a main minimum and two side minima and
the values of the constants $k$'s  in the various contribution to the potential
are proportional to those of Ref.~\cite{mossa07}:
$k_{\rm r}=418.41$, $k_\theta=83.682$,  $k_{\varphi}^{(1)}=4.1841$, $k_{\varphi}^{(2)}=2.092$.

\section{Sampling the landscape}
\label{sec2}

The energy landscape of each polymer has been explored by means of a new take on the activation
relaxation technique (ART)~\cite{barkema96,mousseau98}. The original ART  is a method to jump
from a minimum of the potential energy to a neighboring one, and so on with a sort of random
walk, until the space of minima has been satisfactorily explored. Each local minimum is characterized
by a positive curvature of the energy function in all directions. The matrix with the second partial
derivatives of the potential energy is the Hessian $\caH$, and at a minimum it has all
non-negative eigenvalues. The escape from a minimum with the ART goes via an activation that
attempts to find  first a nearby saddle in the energy landscape. This is done by slowly forcing
the chain in a random direction of the $3N$-dimensional
configuration space until a negative  eigenvalue of
$\cal{H}$ arises. The direction of the negative eigenvalue is then  followed  until the force
vanishes, which indicates that a saddle point  of the energy function has been reached. By a
gentle push to the other side of the saddle and with a subsequent minimization, eventually a new
minimum is reached. The new found minimum is then added to the catalog of minima if not already
present. The same is done for the saddle with a separate catalog.
In our case, configurations of the newly recovered minima and saddles are compared to the
already recorder ones by means of a contact distance analogous to that described in \cite{veit}.

In our model, as in other polymer models where ART has been
used~\cite{wei03,mousseau05,Yun06}, one cannot deform a configuration at random during
activation, because the action of the strong spring force (with constant $k_r\gg k_\theta
\gg k_\varphi^{(1)}$) prevents the system from accurately following the direction of the
negative eigenvalue. Indeed, upon forcing the bond potential $V_{\rm bond}$ with a random
deformation,  the configuration always bounce back to the minimum without reaching any
saddle. We have thus chosen to perturb only one dihedral angle at a time, while the
rest of the molecule is allowed to deform according to the potential in order to find the minimum
energy compatible with the imposed dihedral. When a negative eigenvalue is found, it is
followed with a deformation along the same direction while energy is still minimized
in the orthogonal directions, as in usual ART. This is the point that indeed fails if the
random deformation is performed.

Contrary to usual ART, our approach involves a finite amount of possible deformations,
because only $2(N-3)$ possible changes of dihedral angles can be tried. This feature is not
intended to promote  a random diffusion in the space of minima, as in standard ART
implementations, but rather to implement a systematic protocol for the cataloging of all
saddles and minima. More precisely we adopt the following procedure: starting from an initial
catalog with just one minimum, we try all possible activations from single dihedral
deformations of that configuration. For each transition to another minimum, as before, we
check whether the minimum is already in the catalog and eventually add it (and the same for
the saddle). After all deformations for the first minimum have been attempted, we repeat the
process from the second minimum of the catalog, and so on. The algorithm stops at the $M$'th
minimum if no new minima are found.  At this point the catalogs of minima and saddles are
considered complete. One can view the whole process as an attempt of exact enumeration of
minima and saddles. The choice of a finite set of activation moves, which might in principle
lead to poor sampling of the configuration space, is justified in our case by the limited
number of essential degrees of freedom (the $N-3$ dihedral angles).

Another delicate numerical issue is how to precisely follow the negative eigenvalue
direction: if the negative eigenvalue direction is followed soon after it has been
detected, the configuration could bounce back to the minimum, and consequently one could miss
a saddle. To avoid this problem, the dihedral deformation is continued until the negative
eigenvalue is lower than a small threshold $e<0$. In principle this procedure
might exceedingly deform the configuration and lead to a saddle that does not belong to the
basin of the minimum where one has started from. However, cross checks in the transitions
from and to minima indicate that this effect, if present, is very small.

A substantial portion of the algorithm we used is based on software downloaded from
N.~Mousseau's  web-page \cite{mousseauweb}, version 2006. This
software is portable and allows for an easy replacement of the energy function, which was for
Lennard-Jones clusters of atoms in origin.

\section{Renormalization}
\label{sec3}

\begin{figure}[tb!]
\includegraphics[clip,width=7.95cm]{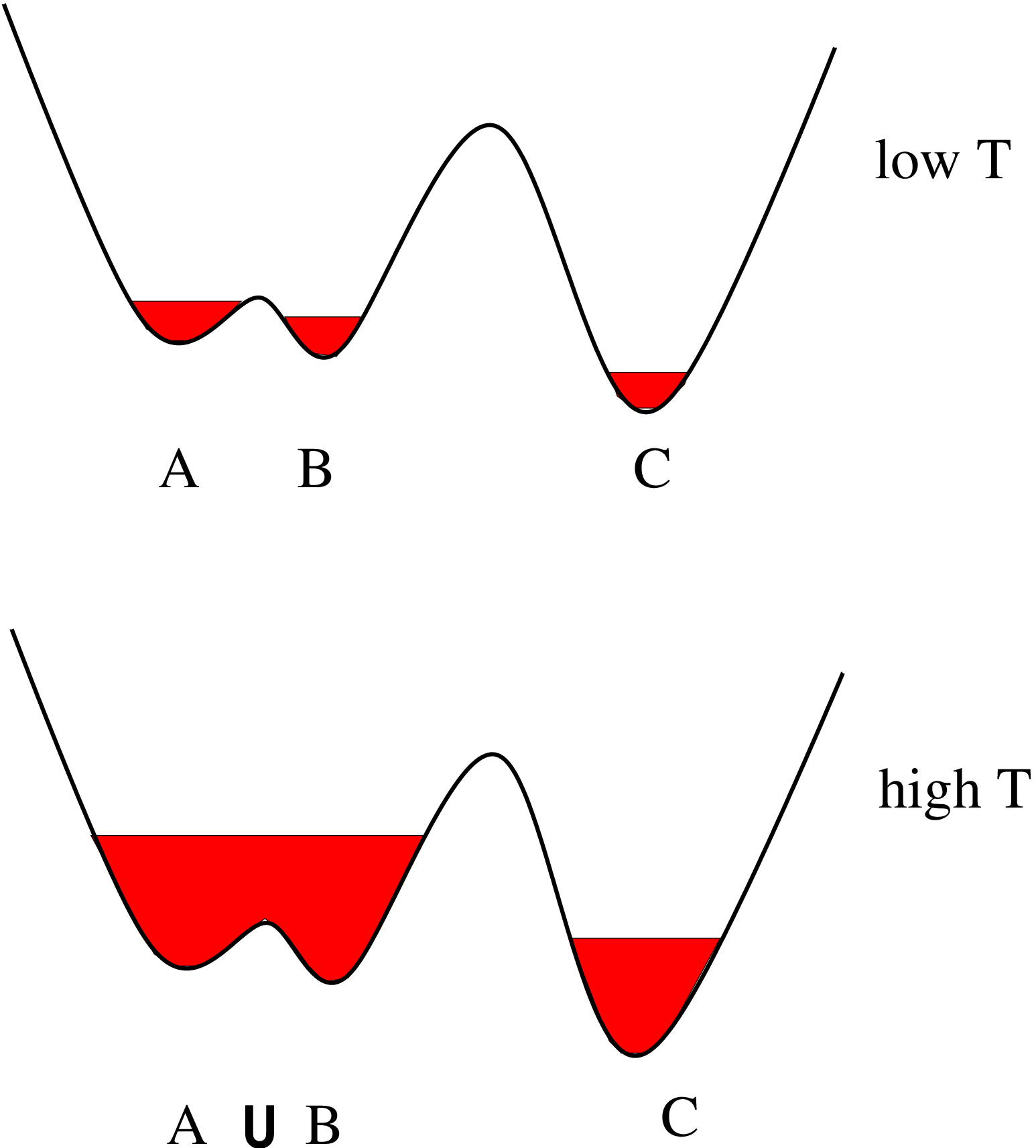}
\caption{Sketch of temperature dependence of the partitioning in metastable states
in a three-well potential model.
At low temperature the saddle separating the minima A and B is
sufficiently high to provide a separation of time scales. When thermal agitation increase, the explored regions around the minima of
the energy landscape (shaded areas) become wider.}
\label{tempchange}
\end{figure}

As already noted in the introduction, the modeling of polymer dynamics as a hopping process relies on the
existence of a separation of time scales. The configuration space of many systems at sufficiently low
temperature, like collapsed globules and glasses, can be partitioned into metastable states, regions
whose internal sampling time  is significantly lower than the escape time. In the traditional double-well
picture it is straightforward to identify such regions as the  basins of attraction of different local minima of
the potential energy, and the crossing of the first-order saddle that separate them as the time-limiting
step in the exploration of the landscape. This picture, however, fails to correctly reproduce the
appropriate division of time scales in more complex potential energy landscapes, with many minima separated
by energy barriers that span a wide range of energies. In these cases many minima of the potential energy
are separated from their neighbors by saddles that are much lower than the available thermal energy and
therefore fail to  represent effective kinetic bottlenecks. In these conditions metastable states do not
consist anymore of single minima of the potential energy, but are instead composed of a collection of
basins of attraction of different minima separated  by small activation energies.
Figure~\ref{tempchange} illustrates how the configuration space partitioning into metastable states depends on
temperature: the shaded areas in the figure represent the regions mostly visited due to thermal agitation,
while the rest of the landscape is only rarely explored. An energy barrier able to
sufficient to dynamically isolate the basins of attraction of two different minima $A$ and $B$ at low temperatures might not
represent a relevant kinetic barrier at higher temperatures anymore. Hence the two minima must be
considered as belonging to the same metastable state $A\cup B$.
Only in the zero-temperature limit a metastable state
corresponds to the basin of attraction of a single local minimum of the potential energy.
In this case the NMS can be easily determined as the graph whose nodes are the potential energy minima
and the links are first-order saddles connecting their basins of attractions.

%%%%%%%%%%%%%%%%%%%%%%%%%%%%%%%%%%%%%%%%%%%%%%%%
%%%%%%%%%%%%%%%%%%%%%%%%%%%%%%%%%%%%%%%%%%%%%%%%

\begin{figure}[tb!]
\includegraphics[clip,width=7.95cm]{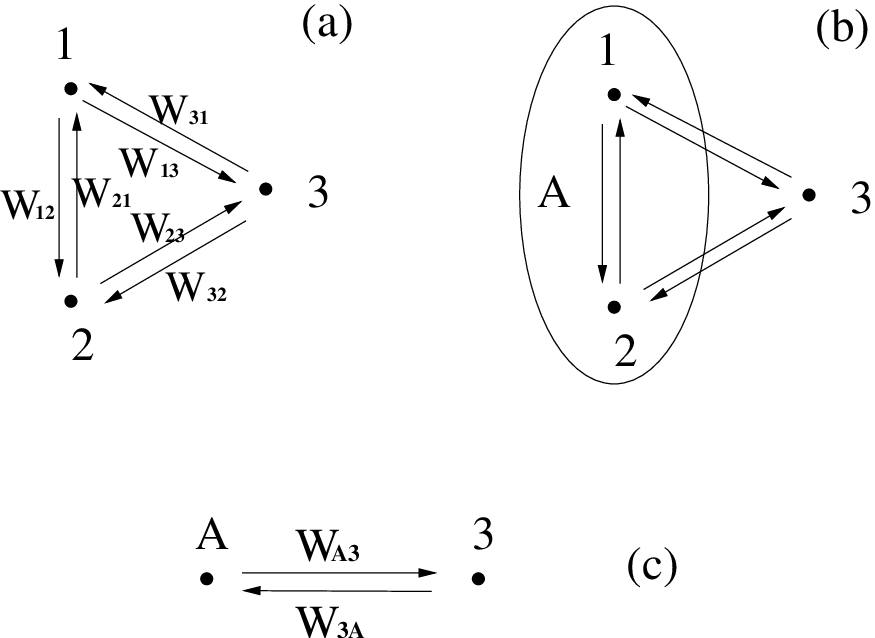}
\caption{Coalescence of two nodes. If one of the energy barriers separating 1 from 2, $W_{12}$
or $W_{21}$, is smaller than the reference temperature, the two nodes coalesce together forming
a new node $A$ that is connected to node 3 by a single energy barrier, which is min$(W_{13},W_{23})$
to go from $A$ to 3, and min$(W_{31},W_{32})$ to come back from 3 to $A$.}
\label{renormalizationsteps}
\end{figure}

In order to determine the NMS at a given non-zero temperature we employ the recursive
renormalization procedure proposed in \cite{grafi2D}. Its main ingredient is the  single
coalescence step illustrated in Fig.~\ref{renormalizationsteps}: if one of the energy
barriers separating node 1 from node 2, $W_{12}$  or $W_{21}$, is smaller than the current
temperature, the two nodes coalesce together forming  a new node $A$. Between the possible
paths from and to the new node, those with minimal energy barrier are kinetically the more
relevant. We chose therefore to consider $A$ as connected to the rest of the NMS only by means
of the minimal energy connections. For example,  considering node 3 in
Fig.~\ref{renormalizationsteps}, the relevant connection would be  $W_{A3}=$min$(W_{13},W_{23})$ and
$W_{3A}=$min$(W_{31},W_{32})$.

The actual implementation of the algorithm is as follows.
First of all, we sort all connections in ascending order according to their energy barrier.
Then, starting from the first connection,
we apply the following iterative procedure to all connections whose  energy barrier is lower than the reference temperature:
\begin{itemize}
\item[$(i)$] The minimal energy node
            connected by the selected saddle is identified and its
            label replaces the label corresponding
         to the other node interested by the connection in the entire connections database.

\item[$(ii)$]  The database of connections is searched for eventual multiple connections between the
            same pairs of nodes. When such connections are found only the one characterized
         by the lowest energy barrier is kept and the other ones are erased.

\item[$(iii)$] The selected connection is finally erased from the database.
\end{itemize}
This renormalization algorithm allows to determine the NMS at any given temperature
starting from the zero-temperature NMS. As already stressed, the surviving nodes do not
represent minima anymore but collections of minima that are best interpreted as metastable
states.

We notice that in general the renormalization process tends to increase the average connectivity  of the NMS
since the new  node emerging from the coalescence of two neighboring nodes inherits all their connections.
This effect, however is partially mitigated by the presence of shared connections between the 
coalescing nodes. Defining the average connectivity $c$ as the average number 
of links per node and calling $k$ the  number of shared connections between two coalescing nodes, 
it can be easily shown  that 
$c$ will decrease upon renormalization only if $k > c - 2$.

The large scale dynamics of the original molecular model is now suitably described by a
diffusion on the NMS. This process is governed by a linear master equation
\begin{equation}
\frac{{d} P_i(t)}{dt} = - W P_i(t)
\label{matrixmast}
\end{equation}
where $P_i(t)$ is the probability of residing on the $i$-th node of the graph at time $t$,
while $W$ is a temperature dependent Laplacian matrix whose elements are determined by the rates
of transition between different nodes,
\begin{equation}
W_{ij}= \delta_{i,j}\sum_{k=1}^{N}  \Gamma_{jk}  - \Gamma_{ji} \,\,\,
\label{defW}
\end{equation}
with $\Gamma_{jk}$ being the probability per unit time of a transition from node $j$ to node $k$.

If the NMS is globally connected, the Laplacian matrix has only one eigenvector with zero
eigenvalue corresponding to the stationary probability distribution on the graph.
Moreover, it can be shown that the Laplacian matrix of the
zero-temperature NMS is
semi-positive-definite, and direct computation show that this feature is conserved by the
renormalization procedure~\cite{grafi2D}. As a consequence, Eq.~(\ref{matrixmast}) describes
the relaxation to equilibrium on the graph.

By dropping the kinetic information $\Gamma_{jk}$ one can define a discrete Laplacian matrix
$\bar{W}$ given by
\begin{equation}
\bar{W}_{ij} = \left\{ \begin{array}{lccr} 1 & & \text{if ~} {\Gamma}_{ij} \not = 0 \\
0 & & \text{if ~} {\Gamma}_{ij}  = 0 \end{array}
\right.
\label{discretelaplacian}
\end{equation}
that can be interpreted as describing the process of relaxation to equilibrium in a
time unit corresponding to the number of jumps between different nodes. We will see later that
the spectral properties of the discrete Laplacian matrix provide useful insights on the topological
properties of the NMS.

%%%%%%%%%%%%%%%%%%%%%%%%%%%%%%%%%%%%%%%%%%%%%%%%
%%%%%%%%%%%%%%%%%%%%%%%%%%%%%%%%%%%%%%%%%%%%%%%%

%%%%%%%%%%%%%%%%%%%%%%%%%%%%%%%%%%%%%%%%%%%%%%%%%%%%%%%%%%%%%%%%%%%%%%%%%%%%%%%%%%%%%%%%%%%%%%%%%%%%%%%%%%%%%
%%%%%%%%%%%%%%%%%%%%%%%%%%%%%%%%%%%%%%%%%%%%%%%%%%%%%%%%%%%%%%%%%%%%%%%%%%%%%%%%%%%%%%%%%%%%%%%%%%%%%%%%%%%%%
%%%%%%%%%%%%%%%%%%%%%%%%%%%%%%%%%%%%%%%%%%%%%%%%%%%%%%%%%%%%%%%%%%%%%%%%%%%%%%%%%%%%%%%%%%%%%%%%%%%%%%%%%%%%%

\section{Thermodynamics}
\label{sec4}

In order to assign a clear quantitative meaning to the temperature scale in our  model we
determine the folding transition and the $\Theta$-transition  temperatures of each
sequence. The use of the term ``transition'' in this context must be clarified. The systems
we study are far from the thermodynamic limit, so that no sharp thermodynamic transition
occurs and a transition temperature is not rigorously defined. Nonetheless, a change in the
thermodynamic behavior does occur in a relatively narrow temperature range, so that it is
customary to refer to a ``$\Theta$-transition temperature'' and a ``folding temperature''
even for short polymers. When the transitional phenomenon under investigation has a sharp
counterpart in the thermodynamic limit (as is the case with the $\Theta$-transition) the
finite-size transition temperature is determined using methods which would give the correct
result in the thermodynamic limit. The case of the folding transition is different because
it does not have a thermodynamic limit counterpart \cite{dill2,cecco}, and
various methods have been proposed to give a definition of a folding temperature, which
typically yield comparable results \cite{tlp}.

For convenience we measure the temperature in multiples of $k_B$, which is then formally set as $k_B=1$.
The folding transition was primarily determined by using the
``50\% criterion'': the system at its folding temperature has equal probability of residing
in the native structure as outside of it.
As far as the $\Theta$-transition temperature is concerned, one should in principle be able to easily
determine it by the presence of maxima of either the specific heat $C_v$
or of the derivative of the gyration radius with temperature, $R'=\partial_T R$.
An estimate of these quantities can be easily computed by
expanding the potential energy at the second order in the fixed points.
The partition function of the system can then be expressed as a sum over all minima:
\begin{equation}
\label{partitionfunction}
Z(\beta)= \sum_i {e^{-\beta V_i} \over \Omega_i}
\end{equation}
where $V_i$ is the potential energy of minimum $i$ and $\Omega_i$ is the product of
non-zero eigenfrequencies of the relative Hessian $\caH_i$.
If we translate this in a local entropy $S_i= - k_B\ln \Omega_i$,
the probability of residing in the $i$-th minimum can be written as a function
of the local free energy $V_i - T S_i$,
\begin{equation}
\label{probability}
p_i(\beta)=\frac{1}{Z(\beta)} {e^{-\beta (V_i -T S_i)}}
\end{equation}
For a given configuration-dependent quantity $\Xi$, the average value on the landscape can then be
expressed as
\begin{equation}
\label{average}
\langle\Xi(\beta)\rangle=\sum_i \Xi_i p_i(\beta).
\end{equation}
In this contest quantities such as the specific heat $C_v=\partial \langle E \rangle / \partial T$
and the average gyration radius (mean square distance of the chain elements form their
center of mass) can be computed once the configurations corresponding to each minimum are known.
The quality of this approximation has been thoroughly tested for
several rough energy landscapes \cite{parisi,wales} showing, as far
as protein models are concerned, a reasonable accuracy at temperatures comparable with the
folding and $\Theta$-transitions \cite{mc}.

\begin{figure}[tb!]
\includegraphics[clip,width=7.95cm]{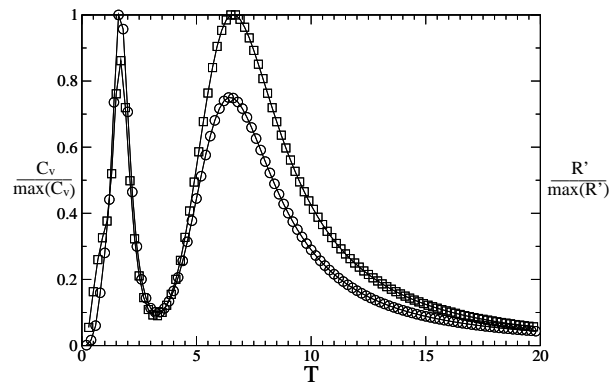}
\caption{Specific heat (circles) and derivative of the gyration radius (squares) as a function of temperature for the
hydrophobic homopolymer of length 12. To ease comparison both quantities are normalized to their maximum value.}
\label{Cv-GyrRad_T}
\end{figure}

\begin{figure}[tb!]
\includegraphics[clip,width=7.95cm]{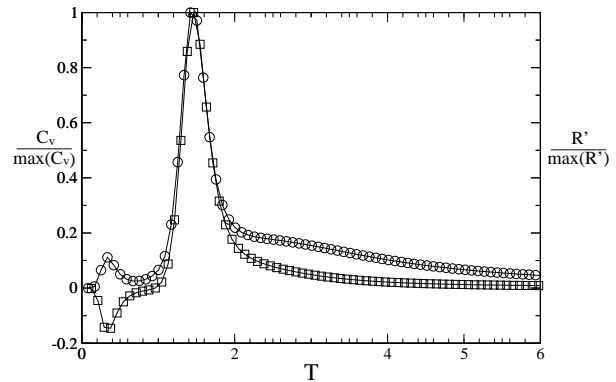}
\caption{Specific heat (circles) and derivative of the gyration radius (squares) as a function of temperature for the
unstable heteropolymer. To ease comparison both quantities are normalized to their maximum value.}
\label{Cv-GyrRad_T2}
\end{figure}

Clearly the definition of the folding transition based on the 50\% criterion suffers of
eventual ambiguities in the determination of the native configuration.
The homopolymers that we model tend to be affected by this problem.
Due to the fact that the energy of homopolymers is invariant upon chain-reversal,
the minima of the potential have a symmetric image, excluding
those for which the two symmetric images coincide.
We find that the probability that the absolute minimum has a non-symmetric
configuration are very high and in the sequences analyzed only the hydrophobic homopolymer
of length $N=10$ does not fall in this category. In all other cases energy has two symmetric global minima,
and they must both be used in order to correctly implement the 50\%
criterion described above.
Analogously, due to the finite size of the systems under study, also the determination of the
$\Theta$-transition temperature might be strongly affected by the parameter used to define it
\cite{tlp}.
In this model the two consensus choices, $C_v$ and $R'$, give very similar
indications. More precisely, for each sequence analyzed, the specific heat shows two
maxima, and the lower one in temperature always coincides with $T_f$ as determined by
means of the 50\% criterion. $R'$~might instead show either
two maxima, or a pronounced minimum and a maximum, the minimum always appearing at lower
temperature than the maximum. The high temperature maximum almost coincide with the
second maximum of the specific heat, thus reinforcing the interpretation
of the latter as a sign of the $\Theta$-transition.
Also the first minimum/maximum always occurs at the same temperature
as the first peak of the specific heat, which, as we already noted, corresponds to $T_f$.
Fig.~\ref{Cv-GyrRad_T} shows this agreement between $C_v$ and $R'$ for the
hydrophobic homopolymer of length 12, while  Fig.~\ref{Cv-GyrRad_T2} shows an example of
negative $R'$ peak at $T_f$ occurring for the unstructured heteropolymer.
%%%%%%%%%%%%%%%%%%%%%%%%%%%%%%%%%%%%%%%%%%%%%%%%
%%%%%%%%%%%%%%%%%%%%%%%%%%%%%%%%%%%%%%%%%%%%%%%%
%%%%%%%%%%%%%%%%%%%%%%%%%%%%%%%%%%%%%%%%%%%%%%%%

An inspection of the gyration radii of individual minima shows that a negative $R'$ peak  appears
when the native minimum does not have the minimal gyration radius.
As a consequence,  the gyration radius might decrease
as soon as the system starts exploring other configurations, eventually increasing again when higher temperatures force it towards swollen conformations.
In both cases the presence of either a minimum or a maximum in $R'$ amounts to a change in
convexity of the $R$ versus temperature and, once again, signals that the system is
undergoing a significant change in its configurational trends.

The $\alpha$-helix-like heteropolymer shows instead only one peak in the specific heat. This thermodynamic
behavior is akin to what observed in similar systems \cite{lapo}, where no molten
globule state could be detected. In those cases $T_\Theta$ and $T_f$ coincide.
Therefore, this sequence is  structurally very stable,
its native configuration being dominant even at temperatures almost as high as the
temperature of thermal unfolding.
Also in this case the derivative of the gyration radius shows a minimum corresponding to the folding
transition, since the native state is an elongated helical structure while other minima have a
smaller gyration radius.

%%%%%%%%%%%%%%%%%%%%%%%%%%%%%%%%%%%%%%%%%%%%%%%%
\begin{figure}[tb!]
\includegraphics[clip,width=7.95cm]{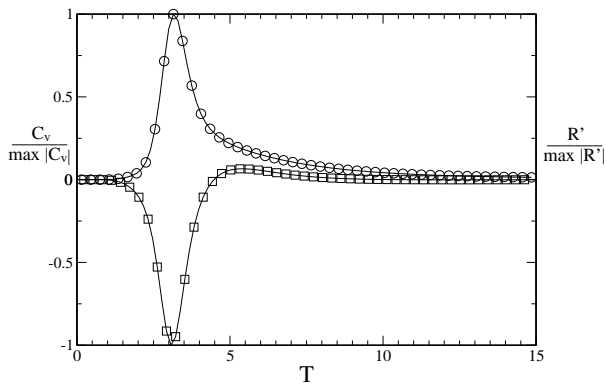}
\caption{Specific heat (circles) and derivative of the gyration radius (squares) as a function of temperature for the
stable heteropolymer. To ease comparison both quantities are normalized to maximum of their absolute
value.}
\label{Cv-GyrRad_T3}
\end{figure}
%%%%%%%%%%%%%%%%%%%%%%%%%%%%%%%%%%%%%%%%%%%%%%%%
In order to illustrate the agreement between the different criteria  for the determination of $T_f$
in table \ref{tempMOD} we report the folding temperature as computed by means of the 50\%
criterion and relying on the first maximum of the specific heat. $T_\Theta$ is also reported.
%%%%%%%%%%%%%%%%%%%%%%%%%%%%%%%%%%%%%%%%%%%%%%%%
\begin{table}[b!]
\begin{tabular}{|c|c|c|c|c|}
\hline
\hfil & \hfil $T_{f,0}$ (n) \hfil & \hfil $T_{f,1}$ \hfil & \hfil $T_\Theta$ \hfil  \\
\hline\hline
\hline
Stable12     &   3.95 (1) & 4.2  &  4.2  \\
Unstable12      &   0.21 (1) & 0.33 &  1.5  \\
Homo12    &   1.25 (2) & 1.6  &  6.5  \\
Homo11    &   1.1  (2) & 1.4  &  4.2  \\
Homo10    &   0.6  (1) & 0.6  &  3.1  \\
Homo09    &   1.3  (2) & 2.0  &  2.5  \\
Homo08    &   1.0  (2) & 1.0  &  0.9  \\
\hline
\end{tabular}
\caption{Folding and $\Theta$-transition temperatures for the sequences under study. Folding
temperatures are determined by means of the 50\% criterion ($T_{f,0}$)
and relying on the first peak of the specific heat ($T_{f,1}$). Near $T_{f,0}$ we report in
brackets the number $n$ of minima used to determine it.}
\label{tempMOD}
\end{table}
%%%%%%%%%%%%%%%%%%%%%%%%%%%%%%%%%%%%%%%%%%%%%%%%
It is interesting to notice that the ratio between $T_f$ and $T_\Theta$, which might be considered as a
relative measure of the stability of their native structure, is approximately the same (4$\div$5)
for the homopolymer of length 12 and the unstable folder.

This table also reveals an interesting feature of the thermodynamics of homopolymers and its dependence on
the system size: while $T_\Theta$ grows with the polymer length (hydrophobic compaction is modelled by two-body
interactions whose effect grows with the chain length), $T_f$ does not show any well defined trend
and oscillates around a fixed value $\bar{T}_f$.
The value of the temperature $\bar{T}_f\simeq 1.1$ determined by using probabilities
is slightly lower than the value $\bar{T}_f\simeq 1.4$ determined
by using the specific heat.

\begin{figure}[tb!]
\includegraphics[clip,width=7.95cm]{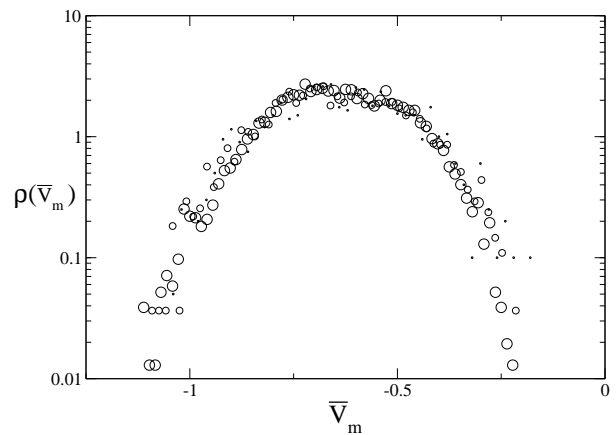}
\caption{Histogram of the rescaled potential $\bar{V}$ of the minima of the potential energy for
the homopolymers of lengths $N=10, 11, 12$. The rescaling factor is $N^{-2}$. The circles radius is
proportional to $N$.}
\label{istoVhomorescales}
\end{figure}

In order to show that
the energy landscape of homopolymers of different length has basically the same shape apart from a scaling
factor $N^2$, in Fig.~\ref{istoVhomorescales} we report the histogram $\rho(\bar{V}_m)$ of the
rescaled potential $\bar{V}=V N^{-2}$ in each minimum. The histograms of the homopolymers of lengths $N=10, 11, 12$
collapse onto the same curve.
It must however be stressed that, since the Lennard-Jones potential is short-range, the number of units that can possibly
interact with a given monomer is limited. As a consequence, for large systems we expect the energy to scale linearly
with the chain length $N$. The almost perfect $\sim N^2$ scaling observed signals therefore that the systems studied are still
very small, their linear dimension being comparable to the Lennard-Jones range.

%%%%%%%%%%%%%%%%%%%%%%%%%%%%%%%%%%%%%%%%%%%%%%%%
%%%%%%%%%%%%%%%%%%%%%%%%%%%%%%%%%%%%%%%%%%%%%%%%

\begin{figure}[tb!]
\includegraphics[clip,width=7.95cm]{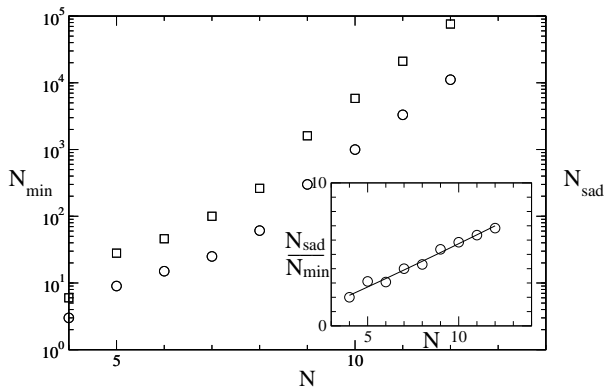}   
\caption{Number of minima $N_{\mathrm{min}}$ (circles) and saddles $N_{\mathrm{sad}}$
(squares) for increasing chain length $N$. In the  inset the ratio
$N_{\mathrm{sad}}/N_{\mathrm{min}}$ is reported together with a power law fit
(exponent 1.07).}
\label{Nminandsad}
\end{figure}

In order to clarify the other observed trend, the independence of $T_f$ on systems size, we
preliminary  observe that increasing the system size leads to a fast increase of the landscape
roughness especially if measured in terms of number of minima and saddles. Figure~\ref{Nminandsad}
shows that, as expected in these systems, $N_{\mathrm{min}}$ grows exponentially with the chain length
$N$. The same holds for the  number of saddles $N_{\mathrm{sad}}$ but with a different
algebraic correction. Indeed, the ratio between these two quantities, which corresponds to half the
average connectivity of the NMS, grows linearly with $N$ (see inset in Fig.~\ref{Nminandsad}).

%%%%%%%%%%%%%%%%%%%%%%%%%%%%%%%%%%%%%%%%%%%%%%%%
%%%%%%%%%%%%%%%%%%%%%%%%%%%%%%%%%%%%%%%%%%%%%%%%

\begin{figure}[tb!]
\includegraphics[clip,width=7.95cm]{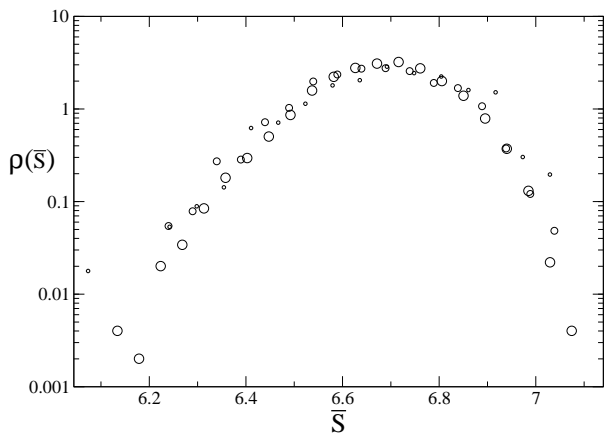}
\caption{Histogram of the rescaled entropy $\bar{S}=S / N^{1.15}$ of the minima of the potential energy for
the homopolymers of lengths $N=10, 11, 12$.  The symbol size is
proportional to $N$.}
\label{istoShomorescales}
\end{figure}

It must now be noted that not only the number of minima increases with system size but also the
total volume of the available configurational space does so. It is thus reasonable to expect that
their ratio, the average volume of the basin of attraction of each minimum, will also experience an
exponential dependence from $N$. The logarithm of the volume of an attraction basin corresponds to
its entropy. As already mentioned, this can be
estimated according to a second order approximation in the local energy minimum:
\begin{equation}
S=-k_B \sum_k^{3N-6}\log(\omega_i^{(k)})
\label{Sclassic}
\end{equation}
where the $\omega_i^{(k)}$'s are the $3N-6$ non-zero eigenfrequencies --- i.e., square root of the
eigenvalues --- of the Hessian matrix in the minimum $i$. In Fig.~\ref{istoShomorescales} we
report the histograms of the entropies for homopolymers of various length after rescaling by a
factor $N^{-1.15}$. The rescaling factor was chosen in order to force the collapse of all
histograms onto a single curve and witnesses   an approximately linear dependence of the average
basin entropy on chain length.
%%%%%%%%%%%%%%%%%%%%%%%%%%%%%%%%%%%%%%%%%%%%%%%%
%%%%%%%%%%%%%%%%%%%%%%%%%%%%%%%%%%%%%%%%%%%%%%%%

As far as the stability of the folding temperature is concerned,
we recall that $T_f$ indicates the point where the free energy of the native configuration
is approached also by other minima. An increase
of the system size would therefore produce two counteracting effects on $T_f$. On the one hand the
average steepness of the landscape will increase, thus increasing the energetic gap between the
native configuration and its neighbors, on the other hand the total number of minima will also
quickly increase, compacting minima in the configuration space and consequently in energy.

\begin{figure}[tb!]
\includegraphics[clip,width=7.95cm]{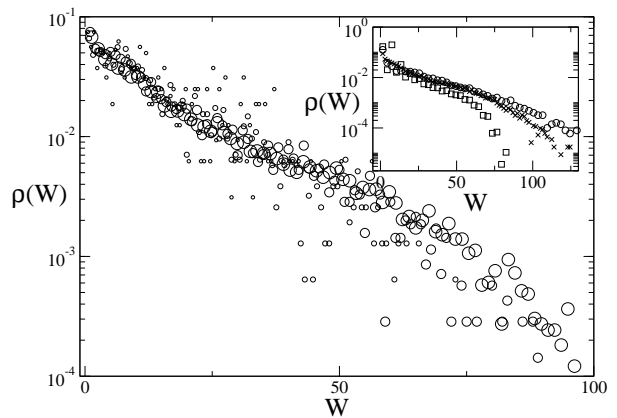}
\caption{Histogram of energy barriers $W$ between connected minima for homopolymers with $N=10, 11, 12$.
The symbol size increases with $N$. Inset: distribution of energy barriers for the unstable folder (squares),
stable folder (circles) and homopolymers with $N=12$ (crosses).}
\label{distribarriersCUMU}
\end{figure}

Finally we note that the distribution  $\rho(V_s)$  of the first-order saddles energies
$V_s$  shows a dependency on system size very similar to that of the energy of the minima (data not
shown). It is therefore tempting to  picture the effect of increasing chain lengths as a simple
isometric stretching of the energy landscape, as could be attained by multiplying energy by a constant
factor. This picture apparently clashes with the observation that the profiles
of the histograms of the energy barrier heights $W$ are pretty similar  for all chain lengths,
and consistent with the same exponential function (see Fig.~\ref{distribarriersCUMU}).
%%%%%%%%%%%%%%%%%%%%%%%%%%%%%%%%%%%%%%%%%%%%%%%%
%%%%%%%%%%%%%%%%%%%%%%%%%%%%%%%%%%%%%%%%%%%%%%%%
Also this effect, however, can be accounted for by the quick growth of competing minima. The appearance of
new minima helps breaking conformational transitions in more sub-steps, thus counteracting the general
quadratic increase of  their energetic cost.

%%%%%%%%%%%%%%%%%%%%%%%%%%%%%%%%%%%%%%%%%%%%%%%%%%%%%%%%%%%%%%%%%%%%%%%%%%%%%%%%%%%%%%%%%%%%%%%%%%%%%%%%%%%%%
%%%%%%%%%%%%%%%%%%%%%%%%%%%%%%%%%%%%%%%%%%%%%%%%%%%%%%%%%%%%%%%%%%%%%%%%%%%%%%%%%%%%%%%%%%%%%%%%%%%%%%%%%%%%%
%%%%%%%%%%%%%%%%%%%%%%%%%%%%%%%%%%%%%%%%%%%%%%%%%%%%%%%%%%%%%%%%%%%%%%%%%%%%%%%%%%%%%%%%%%%%%%%%%%%%%%%%%%%%%

\section{The hydrophobic homopolymers: topology as a function of temperature and system size}
\label{sec5}

The thermodynamic properties analyzed so far only depend on the distribution of the minima of
the  potential energy. In order to account for the trends observed in the folding and
$\Theta$-transition  temperature we were led to analyze in some detail the chain length dependence of the
energy landscape both in terms of minima and connection saddles. Since potential energy minima
correspond to nodes of the zero-temperature NMS and saddles correspond to its links, the results
of  the previous analysis might be rephrased in terms of an exponential growth of the graph
degree with system size and a linear growth  of its average connectivity. We will now extend
this analysis to the finite-temperature NMS. More precisely  we will focus on the
combined effects of temperature and systems size on the topology  of the NMS for homopolymers.

The first quantity affected by the renormalization process is obviously the graph size. The  total number of
nodes for each homopolymer is reported in Table \ref{tabNODES} for various temperatures corresponding to
fixed multiples of $\bar{T}_f$ (the average folding temperature for homopolymers, determined using the
peak  in the specific heat) namely $0.5 \bar{T}_f$, $\bar{T}_f$ and $1.5 \bar{T}_f$. Graph sizes appear to
decay exponentially with temperature, with a decay rate growing approximately linearly with the 
chain length.
%%%%%%%%%%%%%%%%%%%%%%%%%%%%%%%%%%%%%%%%%%%%%%%%
\begin{table}[b!]
\begin{tabular}{|c|c|c|c|c|}
\hline
\hfil $N$\hfil & \hfil $T=0.0$ \hfil & \hfil $T=0.7$ \hfil & \hfil $T=1.4$ \hfil &\hfil $T=2.1$ \hfil  \\
\hline\hline
%Good12 & 4128    & 1327  & 507   &    281 \\
%Bad12    & 13394    & 102    & 78    &     36 \\
12  & 11120 & 5276  & 2684  &    1372 \\
11  & 3313  & 1754  & 976   &     532 \\
10  & 999   & 564   & 356   &     228 \\
09  & 300   & 197   & 159   &     106 \\
08  & 104   &  67   &  51   &      36 \\
\hline
\end{tabular}
\caption{Graph size during renormalization at four different temperatures, for
homopolymers of different lengths.}
\label{tabNODES}
\end{table}
%%%%%%%%%%%%%%%%%%%%%%%%%%%%%%%%%%%%%%%%%%%%%%%%
A similar, yet more irregular, decay  can be observed in the number of links of the renormalized graph 
reported in table \ref{tabCONN}.
%%%%%%%%%%%%%%%%%%%%%%%%%%%%%%%%%%%%%%%%%%%%%%%%
\begin{table}[b!]
\begin{tabular}{|c|c|c|c|c|}
\hline
\hfil$N$\hfil   & \hfil $T=0.0$ \hfil & \hfil $T=0.7$ \hfil & \hfil $T=1.4$ \hfil &\hfil $T=2.1$ \hfil  \\
\hline\hline
%Good12  & 2049  & 3233      & 3658     \\           
%Bad12   &163596 & 202     & 154    & 70    \\
%Good12  &31250  & 7930   & 2194   & 974   \\
12  & 73468  & 50802  & 17374  & 5700  \\       
11  & 20392  & 14974  & 7752   & 2456  \\
10  & 5710   & 4236   & 2920   & 1600  \\
9  & 1562   & 1280   & 1130   &  794  \\
8  & 482    & 364    & 298    &  210  \\
\hline
\end{tabular}
\caption{Total number of connections during renormalization at four different temperatures for
five hydrophobic homopolymers of different lengths.}
\label{tabCONN}
\end{table}
%%%%%%%%%%%%%%%%%%%%%%%%%%%%%%%%%%%%%%%%%%%%%%%%

\begin{figure}[tb!]
\includegraphics[clip,width=7.95cm]{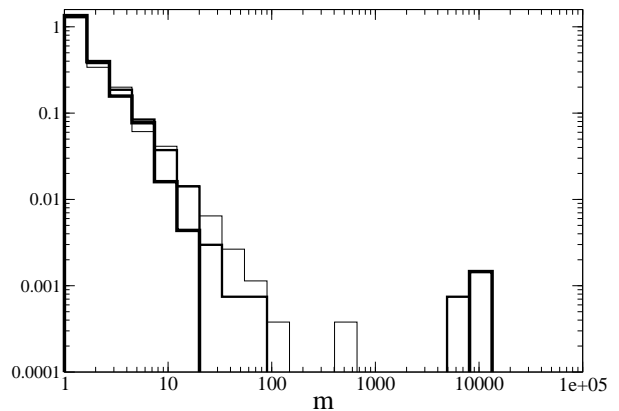}
\caption{Histograms of the multiplicities $m_i$ of each node of the renormalized graph for the
$N=12$ homopolymer at temperatures $T= 0.7, 1.4, 2.1$. Increasing temperatures are represented
by lines of increasing thickness.}
\label{isto-mult-HOMO12}
\end{figure}

In order to understand whether the process of node coalescence that causes the NMS to shrink with
temperature proceeds uniformly on the NMS, we analyze the temperature dependence of the
multiplicities $m_i$ of each node, i.e., the number of nodes of the zero-temperature graph that coalesced into the
$i$-th node. In Fig.~\ref{isto-mult-HOMO12} we report the histograms of $m_i$ for the $N=12$ homopolymer
at various temperatures. The isolated column developing on the right of the figure represents the multiplicity
of the  minimal energy node that is growing at the expense of the nodes with intermediate multiplicity
($m_i>10$). The low multiplicity nodes are almost unaffected. These observations suggest that low saddles
are highly localized around the bottom of the energy landscape. Concerning the latter point, we recall that two nodes
coalesce when they are separated by a link corresponding to a saddle of energy lower than the current
temperature. If the majority of nodes tend to coalesce on the minimal energy node, it means that
this node is, at every temperature, the one characterized by the lowest energy connections. Furthermore,
the fact that isolated nodes of intermediate multiplicity, i.e., isolated regions of the landscape
characterized by the presence of low energy saddles, end up very early in the minimal energy node reinforces
the above picture on the localization of low energy saddles.

The presence of a unique metastable state around and above $T_f$ implies that, at these temperatures, long
homopolymers are characterized by the presence of an extended, fast-connected, low-energy region where
local rearrangements take place more quickly than in the rest of the landscape.

%%%%%%%%%%%%%%%%%%%%%%%%%%%%%%%%%%%%%%%%%%%%%%%%
%%%%%%%%%%%%%%%%%%%%%%%%%%%%%%%%%%%%%%%%%%%%%%%%
The opposite scenario emerges when looking at shorter chains where the fraction of low
multiplicity  nodes decreases. For example in Table \ref{mult1} we report, for the
various  homopolymers, the fraction of nodes with multiplicity 1 at different temperatures.
This is the fraction of nodes still untouched by renormalization and, therefore, separated
from the rest of the NMS by steep activation barriers. Table \ref{mult1} shows that
short homopolymers are more affected by renormalization and loose most of their
nodes of multiplicity 1.
%%%%%%%%%%%%%%%%%%%%%%%%%%%%%%%%%%%%%%%%%%%%%%%%
\begin{table}[b!]
\begin{tabular}{|c|c|c|c|}
\hline
\hfil $N$\hfil  & \hfil  $T=0.7$ \hfil & \hfil $T=1.4$ \hfil &\hfil $T=2.1$ \hfil  \\
\hline\hline
12  & 0.67  &  0.65  & 0.67   \\
11  & 0.67  &  0.63  & 0.65   \\
10  & 0.70  &  0.62  & 0.59   \\
9  & 0.70  &  0.65  & 0.61   \\
8  & 0.65  &  0.52  & 0.41   \\
\hline
\end{tabular}
\caption{Fraction of the nodes of the renormalized NMS having multiplicity
equal to one. Data refer to hydrophobic homopolymers of different lengths at various temperatures.}
\label{mult1}
\end{table}
%%%%%%%%%%%%%%%%%%%%%%%%%%%%%%%%%%%%%%%%%%%%%%%%
This phenomenon might be due to two slightly different mechanisms: either the low energy
node directly attracts low multiplicity nodes or the renormalization process affects in the
same manner nodes of all multiplicities. In both cases the previous argument about the
spatial localization of low energy barriers near the native minimum falls and the landscape
can no longer be divided in  slow and fast regions.
Remarkably, the difference in the spatial distribution of energy barriers on the landscape described above
arises independently from the actual probability distribution of barrier heights
$\rho(V_s)$ (see Fig.~\ref{distribarriersCUMU}) which, we recall, is pretty much size-independent.

Finally, to document the relative importance of the fast-connected central regions in the different
sequences analyzed, we report in Table \ref{tabMULTmax} the number of minima of the potential attracted by
the maximally growing node in the graph. This almost always coincides with the minimal energy node except
for four low temperature cases dubbed with a star in the table. A comparison at the same temperature
between homopolymers of different lengths shows that the fraction of minima attracted by the minimal
energy node (in parenthesis in the table) grows with the chain  length.
%%%%%%%%%%%%%%%%%%%%%%%%%%%%%%%%%%%%%%%%%%%%%%%%
\begin{table}[b!]
\begin{tabular}{|c|c|c|c|}
\hline
\hfil $N$\hfil& \hfil $T=0.7$ \hfil & \hfil $T=1.4$ \hfil &\hfil $T=2.1$ \hfil  \\
\hline\hline
%Good12        & 2049  (0.50) * & 3233 (0.78)           & 3658 (0.89)   \\       
%Bad12         & 13325 (0.995) & 13349  (1)    & 13392 (1)     \\
12     & 453 (0.04)       & 5908  (0.53)  & 8828  (0.79)  \\
11     & 52  (0.02) *    & 1043  (0.31)  & 2324  (0.70)  \\       
10     & 38  (0.04) *    & 117   (0.18)  & 429   (0.43)  \\           
9    &  6   (0.03) *    & 18    (0.06)  & 57    (0.19)  \\       
8    &  4   (0.04) *    & 7     (0.06)  & 21    (0.20)  \\           
\hline
\end{tabular}
\caption{Multiplicity of the maximally growing node during renormalization at three different temperatures for
homopolymers of different lengths. When the maximally growing node does not correspond to the minimal energy node
data are labeled with a star. The fraction of minima coalesced to the minimal energy node is reported
within parentheses.}
\label{tabMULTmax}
\end{table}
%%%%%%%%%%%%%%%%%%%%%%%%%%%%%%%%%%%%%%%%%%%%%%%%
It can be observed that by adding the number of nodes at a given temperature (Table \ref{tabNODES}) to the
number of minima coalesced to the minimal energy node one gets a much higher proportion of the total number
of  minima for long chains than for short ones. For example, at $T=2.1$ the $N=12$ polymer has $8828$ minima in the
minimal energy node and $1372$ nodes surviving on the graph. The latter nodes were much less affected by the
merging process, with  only about a decrease of about 10\% of their number. In the $N=10$
case the decrease is instead of 35\%, once again implying that coalescence is more uniformly  distributed in this
graph.
%_______________________________   fine nodi    _____________________________________________
%_______________________________   Inizio connettivita' _____________________________________________

In the previous section we documented the linear increase of the average connectivity of the graph with the
system size. At finite temperatures this quantity displays a more complex behavior, as shown in Table
\ref{tabCONNmedia} where the average connectivity $c$ of the renormalized  graphs of the five homopolymers is
reported. After an initial growth the longer chains show a decrease of the average connectivity with 
temperature while the two shorter ones exhibit a growing connectivity up to $T=2.1$. 
The temperature of peak connectivity diminishes with the chain length.

\begin{figure}[tb!]
\includegraphics[clip,width=7.95cm]{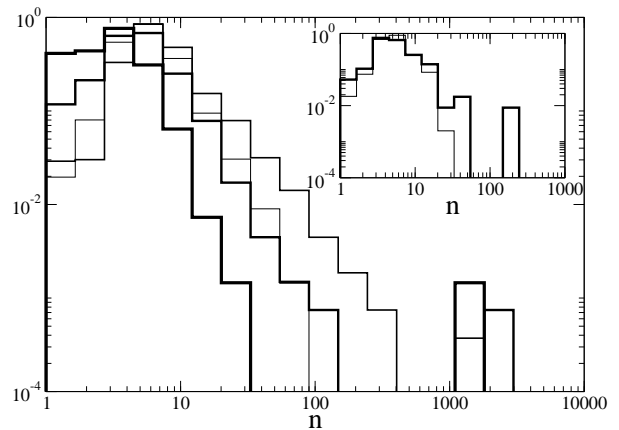}
\caption{Histograms of the number $n_i$ of contacts of each node for the
$N=12$ and $N=10$ (inset) homopolymers at temperatures $T=0.0, 0.7, 1.4, 2.1$ and
$T=0.0, 2.1$ respectively. Increasing temperatures are represented
by lines of increasing thickness.}
\label{istologconnHOMOGLOBALversusT}
\end{figure}

%%%%%%%%%%%%%%%%%%%%%%%%%%%%%%%%%%%%%%%%%%%%%%%%
\begin{table}[b!]
\begin{tabular}{|c|c|c|c|c|}
\hline
\hfil $N$\hfil  & \hfil $T=0$ \hfil & \hfil $T=0.7$ \hfil & \hfil $T=1.4$ \hfil &\hfil $T=2.1$ \hfil  \\
\hline\hline
 12 & 13.2  & 19.3 & 12.9 &  8.4 \\
 11 & 12.2  & 17.0 & 15.9 &  9.2 \\
 10 & 11.4  & 15.0 & 16.4 & 14.0 \\
  9 & 10.4  & 13.0 & 14.2 & 15.0 \\
  8 &  9.2  & 10.8 & 11.6 & 12.0 \\
\hline
\end{tabular}
\caption{The average connectivity $c$ for five hydrophobic homopolymers
of different lengths at various temperatures.}
\label{tabCONNmedia}
\end{table}
%%%%%%%%%%%%%%%%%%%%%%%%%%%%%%%%%%%%%%%%%%%%%%%%
Figure~\ref{istologconnHOMOGLOBALversusT} shows the effect of temperature on the distribution of
the connectivities $n_i$ for the homopolymer of chain length 12 and 10 (inset).
%%%%%%%%%%%%%%%%%%%%%%%%%%%%%%%%%%%%%%%%%%%%%%%%
%%%%%%%%%%%%%%%%%%%%%%%%%%%%%%%%%%%%%%%%%%%%%%%%
While the total number of connections is almost invariant upon renormalization for the shorter systems, 
the longer one shows a 
remarkable loss of connectivity above $n_i=10$. In both cases the growth of an isolated,
high connectivity region is observed. A closer inspection reveals that this region
corresponds to the minimal energy node that, while attracting its neighbors, gradually acquires
new connections. It is thus tempting to ascribe the observed differences in average connectivity
to the different rates of coalescence to this node previously observed in long and short chains.
We remind that the average connectivity $c$ decreases upon renormalization only if 
there are many shared connections between the coalescing nodes. 
The average connectivity  increases significantly during the fast initial growth of the 
``central'' node, signaling that each coalescing node is bringing new
connections. As a consequence the ramifications of the ``central'' node are expanding in areas of the 
NMS which were previously relatively remote in terms of connections. 
After that phase a decrease in connectivity begins, showing that any new node attracted by the fastest 
growing node brings only a few new connections: the central Node has now a 
direct link to almost every portion of the NMS.

%%%%%%%%%%%%%%%%%%%%%%%%%%%%%%%%%%%%%%%%%%%%%%%%
\begin{figure}[tb!]
\includegraphics[clip,width=7.95cm]{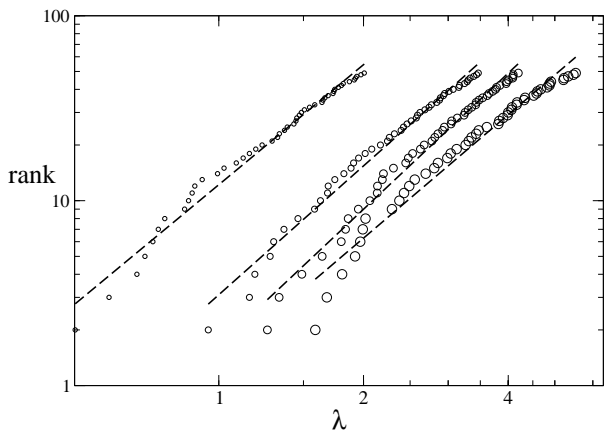}
\caption{Rank-to-eigenvalue plot for the Laplacian matrix of the hydrophobic homopolymer of length 9
at increasing temperature: $T=0.0$, $0.7$, $1.4$, $2.1$. The circles radius is proportional to temperature.
Dashed lines represent least-square power-law fits to the data.}
\label{spectdim09T}
\end{figure}
%%%%%%%%%%%%%%%%%%%%%%%%%%%%%%%%%%%%%%%%%%%%%%%%
%_______________________________    spectra    _____________________________________________

\begin{figure}[tb!]
\includegraphics[clip,width=7.95cm]{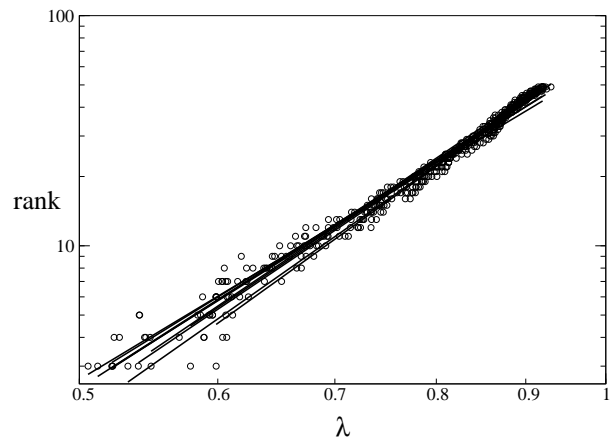}
\caption{Rank-to-eigenvalue plot for the Laplacian matrix of the hydrophobic homopolymer of length 12 for different random
rewirings.}
\label{spectdim12iso}
\end{figure}

We now describe the spectral properties of the NMSs of the sequence studied. More specifically we investigate
their spectral dimension because it determines the large scale diffusivity on the graph. The spectral
dimension describes the spectral density of the discrete Laplacian matrix for small eigenvalues. Since
small eigenvalues correspond to long relaxation paths, high spectral dimensions imply a relaxation to
equilibrium that requires a long series of jumps between nodes.
Hence, although lacking any direct kinetic information,
the spectral dimension provides information about the dynamics of the system.
In Fig.~\ref{spectdim09T} we illustrate the numerical procedure for the determination of the spectral
dimension for the $N=9$ homopolymer at various temperatures. After a numerical diagonalization of the
Laplacian matrix of the NMS the resulting  eigenvalues are rank ordered. The resulting rank-to-eigenvalue
curve is proportional to the integrated  spectral density and can be fitted with a power law in order to
extract the spectral dimension $\bar{d}$. It must be stressed  that one can rigorously assign a topological
meaning to $\bar{d}$ only for asymptotically large graphs: several invariant properties of the
spectral dimension, such as invariance under local link rewiring, only hold in the limit of infinite size.
One might thus try to estimate the effects of the finite size of the analyzed graphs by  quantifying the
amount of such non-invariance. Before evaluating $\bar{d}$ we therefore altered each NMS by adding
random links between second neighbors (non-directly connected nodes connected to each other by a third
node). The procedure was repeated 50 times generating 50 different realizations of each NMS and 50
different estimates of $\bar{d}$ (see Fig.\ \ref{spectdim12iso} where a sample of the rank-to-eigenvalue plots obtained after random
rewirings is reported). Since this quantity should be invariant by link addition the standard
deviation of the values obtained with this procedure provides a measure of the numerical error in its
estimate. The resulting averages $\langle \bar{d} \rangle$ are reported in Table
\ref{tabSPECT}. The associated statistical errors range from 3\% to 5\%.

%%%%%%%%%%%%%%%%%%%%%%%%%%%%%%%%%%%%%%%%%%%%%%%%
\begin{figure}[tb!]
\includegraphics[clip,width=7.95cm]{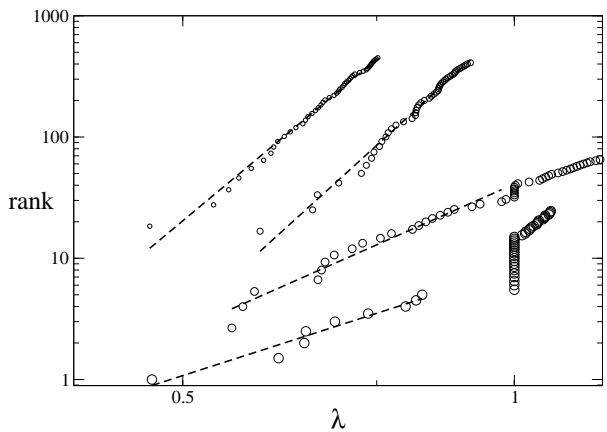}
\caption{Rank-to-eigenvalue plot for the Laplacian matrix of the hydrophobic homopolymer of length 12
at increasing temperature: T=0.0, 0.7, 1.4, 2.1. Larger symbols correspond to higher temperature.
Rank data were multiplied for an arbitrary factor to ease reading. Dashed lines represent least-square
power-law fits to the scale invariant portion of the curves.}
\label{spectdim12}
\end{figure}

%%%%%%%%%%%%%%%%%%%%%%%%%%%%%%%%%%%%%%%%%%%%%%%%
The spectral dimension grows for the homopolymers more quickly than the dimension of the associated
configurational space. On the other hand, at least for short polymers, it does not appreciably change
with temperature coherently with the expected isospectrality of the renormalization procedure.
%%%%%%%%%%%%%%%%%%%%%%%%%%%%%%%%%%%%%%%%%%%%%%%%
\begin{table}[b!]
\begin{tabular}{|c|c|c|c|c|}
\hline
\hfil $N$\hfil & \hfil $T=0.0$ \hfil & \hfil $T=0.7$ \hfil & \hfil $T=1.4$ \hfil &\hfil $T=2.1$ \hfil  \\
\hline\hline
12  &16.4   &16.4   &8*     &5.6*\\
11  &8.8    &12.4*  &*      &*   \\
10  &6.4    &6.0    &5.6    &*   \\
9  &4.2    &4.6    &4.8    &4.4 \\
8  &3.2    &3.0    &2.8    &2.8 \\
\hline
\end{tabular}
\caption{Average spectral dimensions of the five homopolymers for different chain lengths at different
temperatures. Data marked with a star refer to spectra with a significant number of $\lambda$=1 eigenvalues.
Missing data refer to cases where not enough points where available below $\lambda$=1 to allow a reliable
fitting.}
\label{tabSPECT}
\end{table}
%%%%%%%%%%%%%%%%%%%%%%%%%%%%%%%%%%%%%%%%%%%%%%%%
For longer polymers the spectral dimension estimate $\bar{d}$ is affected by an
unexpected spectral feature: at a certain temperature the Laplacian spectral density peaks around
$\lambda=1$.
It can be shown that these eigenvalues are due to nodes connected to more than one leaf node
(by leaf we mean a node with only one link). Initially the presence of a node connected to
two or more leafs is rare but, as renormalization goes on, it tends to increase.
The appearance of $\lambda=1$ eigenvalues is therefore enhanced by a high coalescence rates.
As shown above, long homopolymers are indeed characterized by
the fast coalescing region of the lowest-energy minimum, which explains the early appearance of the mark of
multiple leaf nodes in their spectra.
This phenomenon is depicted in Fig.~\ref{spectdim12} where the rank-to-eigenvalue plot for the
$N=12$ at various temperature is reported. Ranked data have been multiplied by an arbitrary factor to
ease reading.
At low temperatures the curves are pretty much invariant and spectral density is conserved.
At $T=1.4$, however, a step appears in
the curve at $\lambda=1$ (corresponding to a Dirac $\delta$ peak in the spectral density
coming from the abundance of leaf nodes).
Notably a steep drop in the spectral dimension can be observed at the same time and persist at higher temperatures
while the step at $\lambda=1$ grows.
%%%%%%%%%%%%%%%%%%%%%%%%%%%%%%%%%%%%%%%%%%%%%%%%
%%%%%%%%%%%%%%%%%%%%%%%%%%%%%%%%%%%%%%%%%%%%%%%%
Table \ref{tabSPECT}, however, shows that, although the appearance of leaf nodes --- labeled by
stars in the table --- is often associated with changes in the spectral dimension, the latter do not share a
clearly defined direction. In some cases ($N=11$)  $\bar{d}$ increases after the appearance of $\lambda=1$
eigenvalues  while in others ($N=12$) it decreases. Moreover, in a few other cases the spectral
dimension could not be estimated because not enough data where available below $\lambda$=1 (to allow a
reliable fit), and fitting above this value does not seem correct since rank-to-eigenvalue curves
appear either to quickly lose any scale-invariant character for $\lambda=1$ or to display
a markedly different exponent (see the $T=2.1$ curve in fig \ref{spectdim12}).
Such anomalies therefore suggest that the procedure here highlighted for the computation of the spectral
dimension looses meaning in presence of pronounced peaks in the spectral density.

%%%%%%%%%%%%%%%%%%%%%%%%%%%%%%%%%%%%%%%%%%%%%%%%%%%%%%%%%%%%%%%%%%%%%%%%%%%%%%%%%%%%%%%%%%%%%%%%%%%%%%%%%%%%%
%%%%%%%%%%%%%%%%%%%%%%%%%%%%%%%%%%%%%%%%%%%%%%%%%%%%%%%%%%%%%%%%%%%%%%%%%%%%%%%%%%%%%%%%%%%%%%%%%%%%%%%%%%%%%
%%%%%%%%%%%%%%%%%%%%%%%%%%%%%%%%%%%%%%%%%%%%%%%%%%%%%%%%%%%%%%%%%%%%%%%%%%%%%%%%%%%%%%%%%%%%%%%%%%%%%%%%%%%%%
%%%%%%%%%%%%%%%%%%%%%%%%%%%%%%%%%%%%%%%%%%%%%%%%%%%%%%%%%%%%%%%%%%%%%%%%%%%%%%%%%%%%%%%%%%%%%%%%%%%%%%%%%%%%%

\section{The heteropolymers: topology as a function of sequence}
\label{sec6}

The effect of the primary sequences on the topological properties of the NMS is now
determined by investigating the two heteropolymers with 12 residues and 
by comparing the results with those for to the $N=12$ homopolymer.

A first fundamental difference between the stable heteropolymer and the other sequences arises
already at $T=0$ when considering the size of the NMS, which at this temperature simply
corresponds to the number of minima of the potential energy. The strong energetic bias and the
relative lack of frustration of this system result in a much smaller quantity of minima, 4128,
than in the case of the weakly biased unstable heteropolymer which has instead  13394 minima. The
latter is much more similar to the homopolymer which, we recall, has 11120 minima. The lower
number of possible kinetic traps to be overcome by the stable heteropolymer already suggests its
higher folding propensity. On the contrary the picture provided by considering NMS connectivities
is much less clear, since the average number of connecting saddles for each minimum is $16.0$ for the 
stable heteropolymer and $27.4$ for the unstable one while it was $13.2$ for the homopolymer of equal
length. No particular correlation with structural stability can be therefore observed.
%%%%%%%%%%%%%%%%%%%%%%%%%%%%%%%%%%%%%%%%%%%%%%%%
\begin{figure}[tb!]
\includegraphics[clip,width=7.95cm]{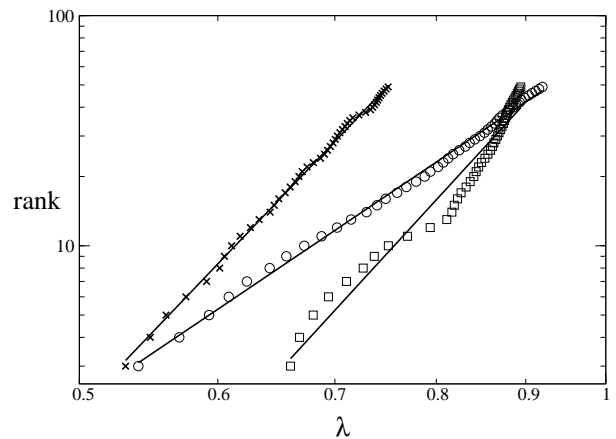}
\caption{Rank-to-eigenvalue plot for the Laplacian matrix of unstable folder (squares), stable folder (circles)
and homopolymer with $N=12$ at $T=0$. Continuous lines represent least-square power-law fits to the data.}
\label{spectdim-sequence}
\end{figure}
%%%%%%%%%%%%%%%%%%%%%%%%%%%%%%%%%%%%%%%%%%%%%%%%

This scenario is further emphasized by a  topological difference between the zero-temperature
NMS of the stable heteropolymer and the other sequences.
While the spectral dimension of the unstable heteropolymer is comparable to that of the homopolymer of
equal length ($\bar{d}=16.6\pm 0.9$ in the first case and $\bar{d}=16.4\pm 0.9$ in the latter), the stable
folder is characterized by a much smaller spectral dimension: $\bar{d}=10.3\pm 0.6$. These differences are
preserved also at higher temperatures since, also in this case, spectral dimensions keep substantially
unchanged until the appearance of $\lambda=1$ eigenvalues. We recall that the unstable  heteropolymer and
the $N=12$ homopolymer also share a very similar ratio $T_f/T_\Theta$. The spectral dimension, a topological
quantity that describes relaxation dynamics, does therefore correlate with the latter ratio, which is a thermodynamical quantity.

This finding relates with analogous evidence \cite{grafi2D} for a two-dimensional toy model where a
difference in the spectral dimension between heteropolymers and homopolymers was detected. In that case, however, the spectral dimension was not found to depend on the stability of the
native structure but rather on the amount of heterogeneity in the sequence. Different sequences displaying
different folding behaviors and landscape steepness but the same ratio of polar to hydrophobic beads shared
the same spectral dimension. Moreover, such a dependency only appeared in that model for finite-temperature
NMSs, while here it is evident already at $T=0$.

The two heteropolymers analyzed here share the same fast decay of the tails of the energy barriers
distribution seen for the homopolymer (see inset in Fig.~\ref{distribarriersCUMU}). In the heteropolymer case
the decay is slightly faster and leads to a substantial cutoff at $V_s\simeq 80$ for the unstable
folder. A far more fundamental difference, however, arises at small energies. The unstable folder
differs from the two other sequences in having a much higher fraction of low  barriers, $V_s<10$.
Since this is the region most affected by the renormalization procedure, it is reasonable to expect
a different behavior of this sequence under renormalization.
Indeed, the comparison of the topological changes induced by the
renormalization process on the NMSs of polymers characterized by
different primary sequences proves very instructive about the origin of the above mentioned
leaf node phenomenon. It is not obvious which physical meaning to assign to
the gradual transformation of the NMS in a {\em stellar} graph
characterized by a center surrounded by leaf nodes. It is also equally
non-intuitive what relation might this process have with the thermodynamics of the
system. The analysis of the spectra of the sequences analyzed (see Table
\ref{flowering}) shows that $\lambda=1$ eigenvalues always become dominant at
$T_\Theta$ implying that at that temperature leaf nodes become the only relevant
feature of the NMS. In fact, at $T_\Theta$ the graph has already
shrunk in size by more than one order of magnitude and almost all of the  missing
nodes appear to have merged to the minimal energy node, 
which seems thus to be the hot-spot of renormalization also for
heteropolymers.

%\ref{tabMULTmax}
%\ref{tabNODES}
%\ref{tabCONN}
%\begin{table}[b!]
%\begin{tabular}{|c|c|c|}
%\hline
%\hfil & \hfil $T=T_\Theta/2$ \hfil & \hfil $T=T_\Theta$ \hfil \\
%\hline\hline
%Good12  & & \\         
%Bad12    & & \\
%Homo12  & & \\         
%\hline
%\end{tabular}
%\label{}
%\caption{.}
%\end{table}

Analyzing the same indicators at lower temperatures (columns referring to
$T=\frac{2 T_\Theta}{3}$ in Table \ref{flowering}) shows, however, a marked
difference in the progress of the appearance of leaf nodes against other
topological transformations of the NMS.
While at $T=\frac{2 T_\Theta}{3}$ the minimal energy node has eaten
up already all the available configuration space, still the relative amount of
$\lambda=1$ is significantly smaller than the value it takes at $T_\Theta$.
The appearance of leaf nodes seems therefore to be intimately connected with
the $\Theta$-transition in these sequences, depicting the moment in which all the
configuration space becomes equally accessible.
In this condition the metastable states still persisting as separate nodes are very unlikely to
have multiple connections because all of their neighbors already collapsed on the minimal
energy node. As a consequence these nodes, which originally corresponded to the frontier of
the low temperature NMS, have a large probability of becoming a leaf node
shortly before coalescing themselves to the minimal energy node.

The pace of this process is strongly sequence-dependent and does not show any clear correlation
with the stability of the native structure. It nonetheless appears to start at markedly
low temperatures for homopolymers.

%%%%%%%%%%%%%%%%%%%%%%%%%%%%%%%%%%%%%%%%%%%%%%%%
\begin{table*}[bt!]
\begin{tabular}{|c|c|c|c|c|c|c|}
\hline
\hfil & \multicolumn{2}{|c|}{surviving} & \multicolumn{2}{|c|}{multiplicity of} &  \multicolumn{2}{|c|}{eigenvalues}\\
\hfil & \multicolumn{2}{|c|}{nodes}     & \multicolumn{2}{|c|}{largest node}           &  \multicolumn{2}{|c|}{with $\lambda=1$}\\
\hline
\hfil & \hfil $T=\frac{2 T_\Theta}{3} $ \hfil & \hfil $T=T_\Theta$ \hfil &
       \hfil $T=\frac{2 T_\Theta}{3} $ \hfil & \hfil $T=T_\Theta$ \hfil &
       \hfil $T=\frac{2 T_\Theta}{3} $ \hfil & \hfil $T=T_\Theta$ \hfil \\
\hline\hline
Stable12 & 155 &  73 & 0.94 &  0.97 & 0.52 &  0.74 \\       
Unstable12  & 402 & 170 & 0.96 &  0.99 & 0.37 &  0.91 \\
Homo12 & 109 &  54 & 0.99 &  0.99 & 0.88 &  0.93 \\       
\hline
\end{tabular}
\caption{Number of surviving nodes, fraction of nodes in the minimal energy node
and fraction of eigenvalues $\lambda=1$ for the three polymers
of length 12 during renormalization.}
\label{flowering}
\end{table*}
%%%%%%%%%%%%%%%%%%%%%%%%%%%%%%%%%%%%%%%%%%%%%%%%

Furthermore, we remember that the the appearance of leaf nodes also depends on
chain length. Shorter polymers have a less marked frequency of $\lambda=1$
eigenvalues also in the proximity of $T_\Theta$
(Table \ref{tabSPECT}). It is therefore worth stressing the differences in
the $\Theta$-transition in short and long polymers. Regardless of chain length,
the $\Theta$-transition signals the temperature at which all configuration space
becomes equally accessible. When this happens in short polymers, the landscape is still
divided into different metastable states or, equivalently, there are still kinetic
barriers to be overcome to access the whole configuration space. On the contrary, in long
polymers, at the $\Theta$-transition almost all the configuration space belongs to the same
stationary state and can be quickly sampled without crossing kinetically relevant barriers.
This does not mean that large barriers are absent from the energy landscape, on the contrary
they are as common as in short polymers. Their arrangement, however, is not capable
of dividing the landscape into kinetically separate metastable states.

%%%%%%%%%%%%%%%%%%%%%%%%%%%%%%%%%%%%%%%%%%%%%%%%%%%%%%%%%%%%%%%%%%%%%%%%%%%%%%%%%%%%%%%%%%%%%%%%%%%%%%%%%%%%%
\section{Conclusions}           

We analyzed the thermodynamics as well as some metric and topological properties of the NMS of short
homopolymers and heteropolymers in a coarse-grained off-lattice protein model. The homopolymers were
hydrophobic in nature and characterized by different lengths, while the two heteropolymers were
chosen in order to ensure a marked difference in structural stability.

The systems investigated exhibit a variety of thermodynamical behaviors with respect to folding.
While  some sequences show all three classical protein conformational arrangements --including
the molten  globule-- others pass directly from folded to swollen, thus reproducing the tendency
of stable folders to have high $T_f$. The model considered is therefore able of efficiently
mimicking the differences of folding propensities observed in real protein sequences.

We have employed a recently proposed procedure \cite{grafi2D} to generate the NMS at a finite temperature $T$,
based on the merging of nodes separated by energy barriers lower than $T$.
This renormalization procedure allows to determine a partition of the configuration space
into dynamically separated regions.
We have thus studied the temperature dependence of the NMS characteristic of each sequence.
The analysis of the topological properties of the NMS of
homopolymers of different lengths highlights how the statistical distribution of activation
energies is not the only factor determining the behavior of the graph under renormalization:
the spatial distribution of the barriers also plays a crucial role.
Indeed, although investigated homopolymers
share the same distribution of energy barriers, their topology respond pretty differently
to an increase in temperature. While in long sequences the renormalization procedure mainly alters the low
energy region of the landscape where metastable states merge into a unique, rapidly expanding
macro-state, in shorter sequences this growth is more  uniformly distributed and other, more
distant metastable states might experience a substantial expansion.
These differences in the renormalization process are reflected in an early (low $T$) tendency to
create leaf nodes in the NMS of long sequences, both for homopolymers and heteropolymers. This phenomenon
is independent on the stability of the native structures of the sequences analyzed and completes
around the $\Theta$-temperature, where the configuration space mostly belongs to same
metastable state.

The spectral dimension of the same sequence at different temperatures is approximately constant,  at
least  before leaf nodes appear. Given the same length, the spectral
dimension appears to be higher for the sequences characterized by unstable native structures. This
finding suggests a non-obvious link between thermodynamics and the dynamical properties of these
systems. Sequences characterized by a stable folded state appear to be characterized by shorter and
less complicated relaxation paths. Indeed, the stable heteropolymer is characterized by simpler
connectivity also in terms of the sheer number of nodes, which, compared to the other $N=12$
sequences, ranges from one half ($T=0$) to one order of magnitude less ($T_\Theta$).

Interestingly some of the features described above seem to hold also
for a simpler two-dimensional model investigated in \cite{grafi2D}. Also in that model the spatial
distribution of the barriers is crucial in shaping some properties of the energy landscape, namely the amount of
kinetic traps. However, in that case only kinetic properties of the NMS were affected and not its
topology. On the contrary, in the model here investigated topology is strongly influenced by the
spatial distribution of energy barriers. Finally, in both models the spectral dimension helps in
categorizing sequences according to the complexity of their relaxation paths. In the two-dimensional
model, however, such a difference only arises at finite temperature and is related to sequence
frustration rather than to the stability of the native structure.

\acknowledgements
We thank C.~Clementi and A.~Mossa for providing us with data.
M.B.~thanks G.T.~Barkema and H.~Vocks for useful discussions,
and acknowledges financial support from K.U.Leuven grant OT/07/034A.
We acknowledge financial support from  EC FP6 project ``EMBIO'' (EC contract n.~012835)

%%%%%%%%%%%%%%%%%%%%%%%%%%%%%%%%%%%%%%%%%%%%%%%%%%%%%%%%%%%%%%%%%%%%%%%%%%%%%%%%%%%%%%%%%%%%%%%%%%%%%%%%%%%%%

\end{document}